Here the paper follows
\documentstyle[preprint,aps]{revtex}

\begin{document}
\preprint{IP/BBSR/94-29}
\begin{titlepage}
\title{Essential Differences Between $ab$ And $c$ Axis \\
 Tunneling And Zero Bias Conductance In The Cuprates.}
\author{P. A. Sreeram$^{*}$ and Manas Sardar$^\dagger$ \\
Institute of physics, Bhubaneswar-751 005, India}

\footnotetext[1]{e-mail:sreeram@iopb.ernet.in}
\footnotetext[2]{e-mail:manas@iopb.ernet.in}
\maketitle
\begin{abstract}
       The peculiarities in tunneling characteristics have been studied
in the light of the controversy between s-wave and d-wave character of
High $T_c$ superconductivity. We show that anisotropic s-wave gap has
the same low voltage power law conductance and two peak structure in
the density of states as d-wave superconductors. The assymetric
tunneling conductance and zero bias conductance for the c-axis
tunneling is shown to occur because of finite band splitting coming from the
interlayer hopping parameter.
\end{abstract}

\end{titlepage}

   Anomolous single particle tunneling characteristics in NIS and
SIS junctions of the high $T_c$ superconductors have remained
a subject of great interest. The single particle tunneling conductance
in both normal and superconducting states is a measure of the
density of states in the normal and superconducting states, and so in principle
one gets a lot of informations about the details of the superconducting gap
parameter. This is  a matter of great current
interest, in view of the recent controversy about d-wave or anisotropic
s-wave symmetry of the gap function.

    Whereas some experimental results ( absence of Hebel-Slichter anomaly
at $T_c$\cite{hebel}, Raman intensity\cite{raman} and low temperature
penetration depth measurements\cite{pene})
are in favour of a d-wave symmetry gap function, the recent experiments
(absence of Josephson current\cite{chou}, temperature
dependent gap anisotropy\cite{jian},
a.c conductivity measurements\cite{ac}) are in favour of an anisotropic
s-wave gap function.

  It has been argued that\cite{shul}, the tunneling characteristics,
specifically the
quadratic rise of current with voltage at low enough voltages and the two
 peak structure seen in the conductance voltage characteristics is an
evidence for a d-wave superconductor. In view of this, we investigated the
single particle tunneling characteristics for the anisotropic
s-wave superconductors , recently proposed by Chakraborty et. al.\cite{cha}

   Before we come to the specific problem we address, we highlight the main
puzzling features seen in the tunneling spectroscopy of the high $T_c$
superconductors\cite{kirt}.

(1) At low temperature, $V=0$ tunneling current is zero for tunneling
along the $ab$ axis, but nonzero along $c$ axis tunneling.
(2) The $ab$ plane conductance becomes smooth at larger temperatures
but the $c$ axis conductance goes on incresing with temperature.
(3) At low temperature and large bias, $ab$ plane tunneling conductance
decreases or saturates whereas the $c$ axis conductance goes on
increasing roughly linearly with voltage.
(4) The $ab$ plane tunneling shows conventional gap like structures but
the $c$ axis tunneling does not show any such charecteristics.
The $c$ axis I-V characteristics shows a much
broadened shoulder at the gap edge.
(5) Both NIS and SIS junction shows assymetric I-V characteristics with
respect to the sign of the bias voltage.
(6) Both direction tunneling shows finite density of states for
$V < \bigtriangleup$
even at the lowest temperatures. For very low voltage current
has a quadratic rise.
(7) There is large broadening of gap in voltage , and a conductance overshoot
for $c$ axis tunneling. Trying to explain this gap broadening due to
inelastic scattering leading to quasiparticle lifetime effect can explain
the gap broadening, but then the zero bias conductance comes out to be
much larger than observed. Invoking strong coupling corrections does not
help, for even though gap broadening of correct magnitude can be obtained,
but it is difficult to explain the conductance overshoot near the gap
edge.
(8) For $ab$ axis tunneling, zero bias conductance is zero. So there is no
density of states at the Fermi energy, but for very
small $V < \bigtriangleup$
 there is finite current, showing that there is no fully developed gap, or the
gap is highly anisotropic with $\bigtriangleup_k$ being very small in a
substantial region of the Brillouin zone.
For the $c$ axis tunneling the most common explanation for the ubiquitous
zero bias conductance and the characteristic $V$ shaped conductance
versus voltage characteristics is explained , as due to either because
of tunneling through localized states in the barrier or due to
scattering by magnetic impurity inside the junctions. The first process
is known to give rise to non trivial energy dependence of the
tunneling probability and can lead to zero bias
conductance, and the second process was invoked by many people to explain
non linear current voltage characteristics in tunnel junctions above the gap.
It is worth emphasizing that these peculiar features of the $c$ axis
tunneling are seen in point contact and break junction measurements also.

  In this paper we emphasize on the distinction between the $ab$ plane
  and $c$ axis single particle tunneling channels for both supeconductor
to normal and superconductor to superconductor (NIS and SIS) junctions.
Specifically we shall consider a layer material
  like YBCO or Bi-2212 material. Generalisation to multilayer
  system is trivial. We model such superconductors by two planar
  BCS superconductors coupled by a single particle hopping term along
  the $c$ axis. We consider also the case, when over and above the
  single particle hopping, there is a Josephson coupling between the planes.
We propose that the observed assymetry of the normal state
in plane tunneling conductance, with respect to the sign of the
bias voltage in MIC(metal-insulator-cuprate) junction is a consequence of
the existence of  nonbonding and bonding band with finite splitting
between them, in the cuprates. So far there is no agreement on this
observed assymetry. Barrier shape effects cannot explain it, because
it requires an unusually low barrier height.
In the split band picture, when the metal is held at positive bias with
respect to the cuprate, then there are two channels of elastic
tunneling into the nonbonding and the bonding bands. In the reverse
bias situation, only one of the bands takes part in tunneling.
So the conductance will be assymetric, for MIC and
NIS(here both sides are cuprates, but one is superconducting
and the other is in the normal state) junctions. On the other
hand for SIS and CIC junctions the conductance voltage characteristics
will be symmetric.

  Now for a CIC , NIS or for a SIS junction, when both sides of the
junctions are cuprates, there is an important difference
between tunneling along $c$ and $ab$ axis. In the $ab$ axis tunneling geometry
electrons tunnel only from antibonding to antibonding and
bonding to bonding bands. Whereas for the $c$ axis tunneling there is
another additional channel for conduction, i.e from
antibonding to bonding band. This tunneling will be present even
in absence of a finite bias voltage either way. The chemical potentials
for the two bands differ by $2t_{\perp}$, where $t_{\perp}$
is the $c$ axis hopping amplitude. So the tunneling along $c$ axis
will show a zero bias conductance, but the $ab$ axis tunneling
will have zero conductance at $V=0$ and $T=0$.

 We find that the zero bias tunneling conductance observed along the
$c$ axis tunneling increases with temperature and do not show any
sign of saturation at all.
  This is our main result.

  We shall also discuss, the reason why for $T > T_c$ the
$ab$ axis tunneling
  characteristics becomes smooth, while the $c$ axis tunneling continues to
  be temperature dependent and rises with temperature. Lastly we
  predict that for MIC geometry tunneling (below $T_c$) the assymetry
  (or alternatively the background conductance) will be more
  for lesser value of the gap in the superconductor.

To start we take the effective hamiltonian proposed by Chakraborty
et. al.\cite{cha}
\begin{eqnarray}
{}\sum_{k}(\epsilon^{1}_{k}-\mu )c_{k\sigma }^{1\dagger }c_{k\sigma }
^{1} ~~&+&~~(1\rightarrow 2)~~+~~V_{bcs}\sum_{kk^{\prime }}c_{k\uparrow}
^{1\dagger}c_{-k\downarrow}^{1\dagger}c_{-k^{\prime}\downarrow}^1
c_{k^{\prime}\uparrow}^1 ~~+~~(1\rightarrow 2)\nonumber \\
{}~~&+&~~\sum_{k}{t_{\perp}^2(k)\over t}c_{k\uparrow}^{1\dagger}
c_{-k\downarrow}^{1\dagger}c_{-k\downarrow}^2c_{k\uparrow}^2.
\end{eqnarray}
In this model, there is no hopping term along the $c$ axis from
plane to plane, even though the band theory estimates for the $c$ axis hopping
amplitude $t_{\perp}$ is about ${1\over 3}$ to ${1\over 5}$
of the inplane hopping parameter. The reason is supposed to be that,
due to strong correlation in the plane itself, the single particle band
motion is absent in the $c$ direction. The conduction along $c$ axis is
purely due to incoherent processes. This phenomenon is termed as ``confinement"
by Anderson. The net effect of being, that there is no need to keep
the single particle hopping term along $c$ axis. On the other hand
coherent propagation of ``singlet objects"( pairs of elctrons)is possible.
That is the origin of the last term ( Josephson coupling of a very
unusual kind). It should be emphasized that, it has not been proved
within a realistic model for hig $T_c$ superconductors.

  We prefer to keep the band term in the hamiltonian. The origin of
subbands can be understood as follows. We consider a two layer material like
YBCO. Th individual layers can be modelled by a 2-d tight binding
band with dispersion,
$$
\epsilon_k~~=~~-2t({\rm cos}(kx)+{\rm cos}(ky)) ~~+~~4t^{\prime}
{\rm cos}(kx){\rm cos}(ky)
$$
where $t$ and $t^{\prime}$ are nearest and next to nearest neighbour
hopping in the planes of some effective site.
We take, $t=0.3{\rm eV}$ and $4t_{\perp}=0.45 {\rm eV}$.
For two closely spaced planes, in interlayer matrix element
$$
t_{\perp}(k)~~=~~t_{\perp}({\rm cos}(kx)-{\rm cos}(ky))^2
$$
results in formation of subbands,
$$
E_{\phi,\psi}(k)~~=~~\epsilon (k)\pm t_{\perp}(k)
$$
Where the $\phi$ and $\psi$ are antibonding and bonding band fermions
defined as,
$$
\phi(k),\psi(k)~~=~~{(c_{k}^1\pm c_k^2)\over 2}
$$
The location of chemical potentials will be detrmined by the
doping.

     To illustrate the difference between tunneling along $ab$ axis
and along $c$ axis we write down the tunneling hamiltonian without
any explicit dependence of the tunneling amplitude on momenta or
energy. For tunneling along $ab$ axis, the hamiltonoan will be
$$
\sum{kp}T_{kp}c_{p\sigma}^{\alpha}c_{k\sigma}^{\alpha} \equiv
{}~~\sum_{kp}T_{kp}(\phi_{k\sigma}^{\dagger}\phi_{k\sigma} ~~+~\phi\rightarrow
\psi)
$$
where $\alpha$ denotes layer index (1 and 2).
The $c$ axis tunneling hamiltonian on the other hand will be
$$
\sum{kp}T_{kp}c_{p\sigma}^{1}c_{k\sigma}^{2}~~+~~1\rightarrow 2 \equiv
{}~~\sum_{kp}T_{kp}(\phi_{k\sigma}^{\dagger}\phi_{k\sigma} ~~+~\phi\rightarrow
\psi ~~+~~
{}~~\sum_{kp}T_{kp}(\phi_{k\sigma}^{\dagger}\psi_{k\sigma}~+~h.c)
$$
It is clear, that for $c$ axis tunneling there is an extra channel
for conduction , i.e from the nonbonding to bonding band which is
absent for the $ab$ axis tunneling.

{\it MIC junction tunneling }.

 To illustrate the effect of this band splitting in the normal
state itself , let us consider the MIC junction
tunneling. Within the independent
electron approximation, the single particle tunneling current is given by,
$$
I~~\equiv~~ \int \vert T\vert^2 ~~N_1(E)N_2(E+eV)\lbrack f(E)-f(E+eV)
\rbrack~dE
$$
where energy $E$ is measured from chemical potential,$N_1$ and $N_2$
are density of states of two electrodes (one usual metal and the
other one being the cuprate ), $V$ is voltage bias and $f$ is the Fermi
function. If we take the density of states of the metal $n_2$=const and
that of the cuprate to be $N_1(E)=N_{\phi}+N_{\psi}\theta(-2t_{\perp}-E)$
where $N_{\phi}$ and $N_{\psi}$ can be taken as constants for
simplicity,then
from  the above two equations , we get,
$$
{dI\over dV}~\approx ~\vert T\vert^2 N_{\phi}\lbrack 1+\alpha f({-2t_{\perp}
-V\over k_BT})\rbrack
$$
with $\alpha=N_{\phi}/N_{\psi}$. For $T=0$, $G(V)\equiv 1+\alpha$.
  It is clear that, the tunneling conductance is assymetric with respect to
bias. For positive bias, $G(V)~\approx ~\vert T\vert^2 N_{\phi}(1+\lambda)$
and for negative bias, $~\approx ~\vert T\vert^2 N_{\phi}$ if $\vert V \vert
> 2t_{\perp}$ and $\approx \vert V\vert^2 N_{\phi}(1+\lambda)$ otherwise.
This is true at $T=0$. At nonzero temperatures , the conductivity assymetry
will be seen at lower bias and the absolute value of conductivity
will decrease.

At this point, we compare our model with that of Levin and Quader\cite{levin},
who also consider a split band picture.  We insist that there is a
major difference between our viewpoints as regards the role of the
split bands. Levin et. al.\cite{levin} assume that the bonding
band($\psi$ band)
is almost submerged below the Fermi surface. For the underdoped case,
the $\psi$ hole band is completely filled and frozen much below the
Fermi surface, and do not take part in tunneling to the metal on
the other side of the junction. Consequently there will not be
any conductivity assymetry for underdoped case. For larger doping
case, both the bands will be partially filled and take part in
tunneling. Moreover one needs additional assumptions, that the $\psi$
band is actually a band of nondegenerate band of fermions since their
number is so small. One needs to have, in an adhoc fashion, different
dispersion for $\phi$ and $\psi$ fermions(linear and quadratic in
momenta) to reproduce some normal state properties.
This picture  is approximately right when $t_{\perp}$ is large,
giving rise to large band splitting. We assume , on the other hand
that the band splitting is small (small $t_{\perp}$). So, even at small doping
concetrations , both bands will be partially filled.

Within the interlayer tunneling mechanism of superconductivity,
even though the intralayer BCS coupling gives a small $T_c\equiv 5 K$
on its own, a very small $t_{\perp}$ is enough to raise the $T_c$ to
large values $\equiv 90 K$ through the Josephson coupling term.
We have not yet made a detailed study of the doping dependence on the
MIC tunneling.
In other words a small band splitting explains the observed
assymetry in tunneling conductance at small doping as well as very
high $T_c$ in these materials.

{\it CIC junction tunneling}

 For CIC junctions when both the electrodes are high $T_c$ materails
(break junctions), a look at the tunneling hamiltonians for the
$ab$ and $c$ axis tunneling shows that, for $ab$ tunneling, the elctrons
tunnel from $\phi$ to $\phi$ and from $\psi$ to $\psi$ bands only. For the
$c$ axis tunneling, cross tunneling also takes place. If $T_{\phi \phi}$
$T_{\psi \psi}$ and $T_{\phi \psi}$ are the tuneling matrix elements
between the respective subbands of both electrodes, we get
$$
G_{ab}(V)~~=~~\vert T_{\phi \phi}\vert^2N_{\phi}^2\lbrack(1+\alpha_1^2)
{}~-~\alpha_1^2~f({-2t_{\perp}-V\over k_BT})\rbrack
$$
and
$$
G_c(V)~~=~~\vert T_{\phi \phi}\vert ^2N_{\phi}^2\lbrack (1+\alpha_1^2
+2\alpha_2)~-~\alpha_2~f({-2t_{\perp}+V\over k_BT})~-~(\alpha_2+\alpha_1^2)
{}~f({-2t_{\perp}+V\over k_BT})\rbrack
$$
where $\alpha_1^2=N_{\psi}^2T_{\psi \psi}^2/N_{\phi}^2T_{\phi \phi}^2$ and
$\alpha_2=N_{\psi}T_{\phi \psi}^2/N_{\psi}T_{\psi \psi}^2$.
 The main features of this expression are:
(1) The conductance voltage characteristics is symmetric with respect
to bias for both $ab$ and $c$ axis tunneling.
(2) For $c$ axis tunneling, there is a zero bias current coming from
cross tunneling, which is operative even at zero bias because of finite
band splitting. For $ab$ axis tunneling there is no zero bias current.
(3) There is a zero bias conductance for both $ab$ and $c$ axis tunneling
At $T=0$ and $V=0$, $G(V)\equiv T_{\phi \phi}^2N_{\phi}^2(1+\alpha_1^2)$.

If we take the tunneling matrix element $\vert T\vert^2 \equiv 1+(V/V_c)^2$,
then the conductivity increses with voltage, but with different slopes
for positive and negative bias for the MIC junctions and with
same slope for CIC junctions. The $1+(V/V_c)^2$ dependence of $T^2$
comes because of Coulomb blockade effects in the junctions. One can also
get a linear conductance for small voltages due to inelastic scattering in the
junctions as we mentioned earlier.
These kind of approaches are specially tailor made to explain the
linear conductance in the cuprates. As we pointed out that the linear
conductance is observed only for the $c$axis tunneling, one has
to explain why, inelastic scattering and coulomb blockade effects are
not seen for the $ab$ axis tunneling also. Moreover the ubiquitous
linear conductance is seen in point contact tunneling also.
We do not attempt to
explain this important feature here. The main thrust of our arguments
is to show the natural origin of the tunneling assymetry in the high $T_c$
materials.
One more important consequence of our model is that
 in the superconducting state,  there is a finite
zero bias conductance
for $c$ axis
tunneling. For the inplane tunneling this is absent.
This will be explored next in SIS and NIS junction tunneling geometries.

{\it SIS and NIS junction tunneling}

The  mean field hamiltonian  in the superconducting phase is,
\begin{equation}
\sum_{k}(\epsilon_{k}+t_{\perp})\phi_{k\sigma}^{\dagger}\phi_{k\sigma} +
\sum_{k}(\epsilon_{k}-t_{\perp})\psi_{k\sigma}^{\dagger}\psi_{k\sigma} \\
+ (V+{t_{\perp}^2\over t})\sum_{k} [(\bigtriangleup^{\star} \phi_{-k\downarrow}
\phi_{k\uparrow} +
\bigtriangleup \phi_{k\uparrow}^{\dagger}
\phi_{-k\downarrow}^{\dagger} ) + \phi \rightarrow \psi]
\end{equation}
 The hamiltonian looks like a sum of two BCS reduced
hamiltonins for the bonding and antibonding electron systems. The generalised
gap equation will be
\begin{equation}
{1\over (V+{t_{\perp}^2\over t})}~~=~~ {1\over 2}\sum_{k} {tanh( \beta E_{k}
^{\phi}/2) \over 2 E_{k}^{\phi}} ~~+ ~~
{1\over 2}\sum_{k} {tanh( \beta E_{k}
^{\psi}/2) \over 2 E_{k}^{\psi}}
\end{equation}
where,
$$
E_{k}^{\phi ,\psi} ~~=~~\sqrt {(\epsilon _{k}\pm t_{\perp})^2 +
\bigtriangleup ^2 }
$$
Going from summation to integral and converting to energy variables
it is not very difficult to see that the $T_c$ is given by

\begin{equation}
k_BT_c~~=~~\sqrt { \omega_{c}^2-t_{\perp}^2} {2e^{\gamma}\over \pi}
e^{- {1\over N(0) (V+t_{\perp}^2/t)}}
\end{equation}

for small values of $t_{\perp}$,
where $e^{\gamma} =1.781$.

 We have solved the gap equation numerically for different
temperatures. In NIS junctions, the tunneling current is given by,
$$
I_{NIS}~=~\equiv \sum_{kp}\vert T\vert^2\lbrack u_k^2\delta (eV+E_k
-\xi_p)\lbrack f(E_k)-f(\xi_p)\rbrack ~~+~~v_k^2\delta (eV-E_k-\xi_p)
\lbrack 1-f(E_k)-f(\xi_p)\rbrack \rbrack
$$
For the SIS junction the corresponding expression is,
\begin{eqnarray}
I_{SIS}~& = &~\sum_{kp}\vert T\vert^2 \lbrack (1-f(E_k)-f(E_p))(v_k^2u_p^2
\delta (eV-E_p -E_k) -u_k^2v_p^2\delta
(eV+E_p+E_k))~~ \nonumber \\
& &
+~~(f(_k)-f(E_p))(u_k^2u_p^2\delta (eV+E_k -E_p)
-vk^2v_p^2\delta (eV+E_p -E_k))\rbrack
\end{eqnarray}
The normalised conuctance versus voltage for the $ab$ and $c$ axis tunneling
are plotted in Fig.1 and Fig.2 respectively. The notable features are,
(1) At $T=0$ there is a sharp voltage threshold for conductivity for
the $ab$ axis tunneling, whereas there is a finite zero bias conductance
for the $c$ axis tunneling.
(2) The sharp voltage threshold for $ab$ axis tunneling gets washed
out at a very small temperature ($4 K$).
(3) The plots for conductance at $T=4,10$ clearly shows the characteristic
two peak structures seen in experiments. For d-wave superconductors
also one gets similar two peak structures.

In Fig.3 we plotted the current versus temperature for $ab$ axis
tunneling for 2 and 5 degree Kelvin. We emphasize that, even
at very low temperatures ( 5 K) the current rises quadratically
with voltage at very low voltages. This clearly shows that, the gap
at most places of the Brilloiun zone is very small and falls faster
with temperatures,
than usual BCS temperature dependence of gap. Thus at small but finite
temperatures  the anisotropic s-wave superconductor becomes indistinguishable
from a superconductor with gap nodes.

 Fig.4 shows the temperature dependence of the  normalised zero bias
conductance
for both $ab$ and $c$ axis tunneling.
 One extraordinary feature of the interlayer tunneling gap function
is that the gap along $\Gamma -M$ direction is large and almost temperture
independent and retains its full gap value at $T=0$ up to about
90$\%$ of the $T_c$, and then falls almost like a weak first
order transition. On the other hand gap in any other direction falls
much faster that the usual BCS gap suppresion due to thermal fluctuations.
The momenta averaged gap also falls very slowly with temperatures,
as observed in the recent photoemission experiments.
This is true when $T_J > V_{bcs}$, or in other words,
when interlayer tunneling is
dominant for very strongly coupled layers.

For weaker $T_J$ or
 with larger in plane $V_{bcs}$, the averaged gap falls faster with
temperatures and slowly approach the usual BCS temperature dependence.
Notice that all these pecularities are only because of the $1-k$ summation
in the interlayer josephson coupling term, as emphasized by Anderson.
For a more conventional josephson coupling,
where the individual momenta of the partners of the cooper pairs
are not conserved, and only the center of mass momenta is
conserved, i.e a josephson term with double momenta summation,
we do not get the above mentioned features at all.

Two things follows automatically  from above discussion.
One is that, in the interlayer mechanism, the gap magnitude in most part
of the BZ is very low (1-3 meV) and also very fragile as far as
thermal fluctuation is concerned. The gap in these regions falls faster than in
the usual BCS gap. This would mean that we shall not get any sharp gap features
at all at any finite temperatures in tunneling experiments. This is what is
observed in our numerical calculations at finite temperatures. In Fig. 3. we
show the $I-V$ characteristics at $T=0$ and $T=5$ degrees for
tunneling along the $ab$ plane.
We see clearly that already at $T=5$ degrees there is finite current
at very low voltages. In other words, indeed it will be very difficult to
distinguish between, the situation where the gap function has
gap nodes on the Fermi surface like in d-wave superconductors, and
the interlayer case.

Note that for tunneling along, the $c$ axis there
will be a finite current for arbitrarily small voltages. So in ceramic
materials, where we measure some average current along both
directions, we shall always get an I-V characteristics looking
just like a superconductors with gap nodes on the Fermi
surface even at $T=0$. For single crystal measurements , and for
 $ab$ plane tunneling there will be a sharp voltage threshold, but
no sharp threshold for small but finite temperatures. It is worth
emphasizing that, we do not really know how impurities and inhomogeneities
suppress the interlayer tunneling gap.

%

{\bf Figure Captions}\\
\begin{enumerate}

\item
 Conductance ( Normalised  with respect to normal state) vs Voltage for
$ab$ axis tunneling for Temperatures 2 (Solid line), 4 (dashed line)
and 10 (dotted line) degrees.
\item
 Conductance ( Normalised  with respect to normal state) vs Voltage for
$c$ axis tunneling for Temperatures 2 (Solid line), 4 (dashed line)
and 10 (dotted line) degrees.
\item
 Current Vs Voltage for the $ab$ axis tunneling, for temperatures
 2 (solid line) and 5 (dashed line) degrees.
\item
Zero Bias Conductance vs Temperature for $ab$ axis (solid line) and
for $c$ axis (dashed line).
\end{enumerate}

\end{document}